# Cryo-Electron Ptychography: Applications and Potential in Biological Characterisation

Chen Huang[1,3], Judy S. Kim[1,2], Angus I. Kirkland[1,2]
1. Rosalind Franklin Institute, Harwell Science and Innovation Campus, Didcot, OX11 0QX, United Kingdom
2. Department of Materials, University of Oxford, Oxford, OX1 3PH, United Kingdom
3. Corresponding author: chen.huang@rfi.ac.uk

## Abstract

There is a clear need for developments in characterisation techniques that provide detailed information about structure-function relationships in biology. Using electron microscopy to achieve high resolution while maintaining a broad field of view remains a challenge, particularly for radiation sensitive specimens where the signal-to-noise ratio required to maintain structural integrity is limited by low electron fluence. In this review, we explore the potential of cryogenic electron ptychography as an alternative method for characterisation of biological systems under low fluence conditions. Using this method with increased information content from multiple sampled regions of interest, potentially allows 3D reconstruction with far fewer particles than required in conventional cryo-electron microscopy. This is important for achieving higher resolution for systems where distributions of homogeneous single particles are difficult to obtain. We discuss the progress, limitations and potential areas for future development of this approach for both single particle analysis and in applications to heterogeneous large objects.

## Introduction

High spatial resolution observation of biological structures in their native state is crucial for understanding biological processes. Transmission electron microscopy (TEM) is an essential tool for achieving this goal, but it is not without its challenges. However, limiting factors, such as low image contrast, electron beam damage, and the need for a broader biological context, can create inevitable compromises.

Progress in structural biology has often been driven by technological breakthroughs in imaging methods. The success of single particle analysis (SPA) [1] is an example of the fusion of cryogenic electron microscopy and structure reconstruction through computational averaging of structural motifs in large datasets. Similarly, cryo-electron tomography (cryo-ET) is routinely used with modern focused ion beam (FIB) milling instruments that generate thin lamellae from frozen hydrated specimens [2].

In recent years, cryo-electron ptychography (cryo-EPty) has emerged as an addition to the biological imaging toolbox. This approach has the potential for requiring fewer particles for SPA due to spatial frequency dependent information transfer that can be controlled through the probe convergence angle [3] and decouples resolution and field of view for a given detector array and pixel size. However there remain challenges in utilising variable higher convergence angles to record data at electron optically limited resolution due to the low contrast of these reconstructions and a hybrid approach using low convergence angle (low spatial frequency) data for particle picking may be needed in the future. At this time the required workflow is far from fully optimised to bring the ease of use of cryo-EPty close to that of conventional cryo TEM [3,4] and the resolution achieved in the physical sciences [5]. We discuss emerging directions and the use of methods developed for cryo-TEM to obtain higher resolution in homogeneous systems.

## STEM and Electron ptychography

Initially, it is helpful to examine similarities and differences to transmission electron microscopy (TEM) and scanning transmission electron microscopy (STEM) (Figure 1). STEM uses a spatially defined probe which is scanned across the sample in contrast to TEM which uses fixed broad beam illumination [6]. In STEM, control of the probe size and convergence angle allow a variety of signals to be recorded, traditionally on monolithic detectors with annular geometries. The recent development of pixelated detectors [7] allows the simultaneous acquisition of signals at different detection angles and subsequent offline synthesis of bright field and dark field images in what is generally referred to as 4D-STEM [8]. In the life sciences STEM has been less frequently used than conventional phase contrast TEM but we note the early cryo-STEM work of Elbaum and Wolf [9,10] and that of Engel and Wall who used STEM for mass thickness determination [11]. More recently the integrated differential phase contrast (iDPC) signal [12] has been used in single-particle cryo-EM analysis below 4 Å resolution. This signal can be acquired using a data acquisition geometry which is almost identical to that reported for electron ptychography but with a differing detector integration and different information transfer. Electron ptychography recovers the complex electron wave function at the exit plane of a sample. The optical geometry used in ptychography is similar to that used in STEM and scans a focused or defocused electron probe across a specimen over a given region of interest (ROI) while maintaining a constant electron fluence and flux comparable to those used in cryo-EM. Convergent beam electron diffraction (CBED) patterns are collected in the far field (Fraunhofer plane) at each probe position, generally using a pixelated detector [13]. This is distinct from diffractive imaging which utilises a single diffraction pattern and a support constraint, thus not generating interference patterns between CBED patterns and not taking advantage of the direct solution that is derived from the Fourier shift theorem [14]. In ptychography, the electron wavefunction at the exit surface of the sample is then computationally retrieved using the redundancy generated by overlapping the coherent illumination in a defocussed probe [8,14,15]. The phase of the wave function is sensitive to high spatial frequency structural details. As an illustration, electron ptychography has surpassed the resolution of state-of-the-art TEM for radiation-resistant samples [5] and has been applied at low fluence to cryogenic biological specimens [3,16].

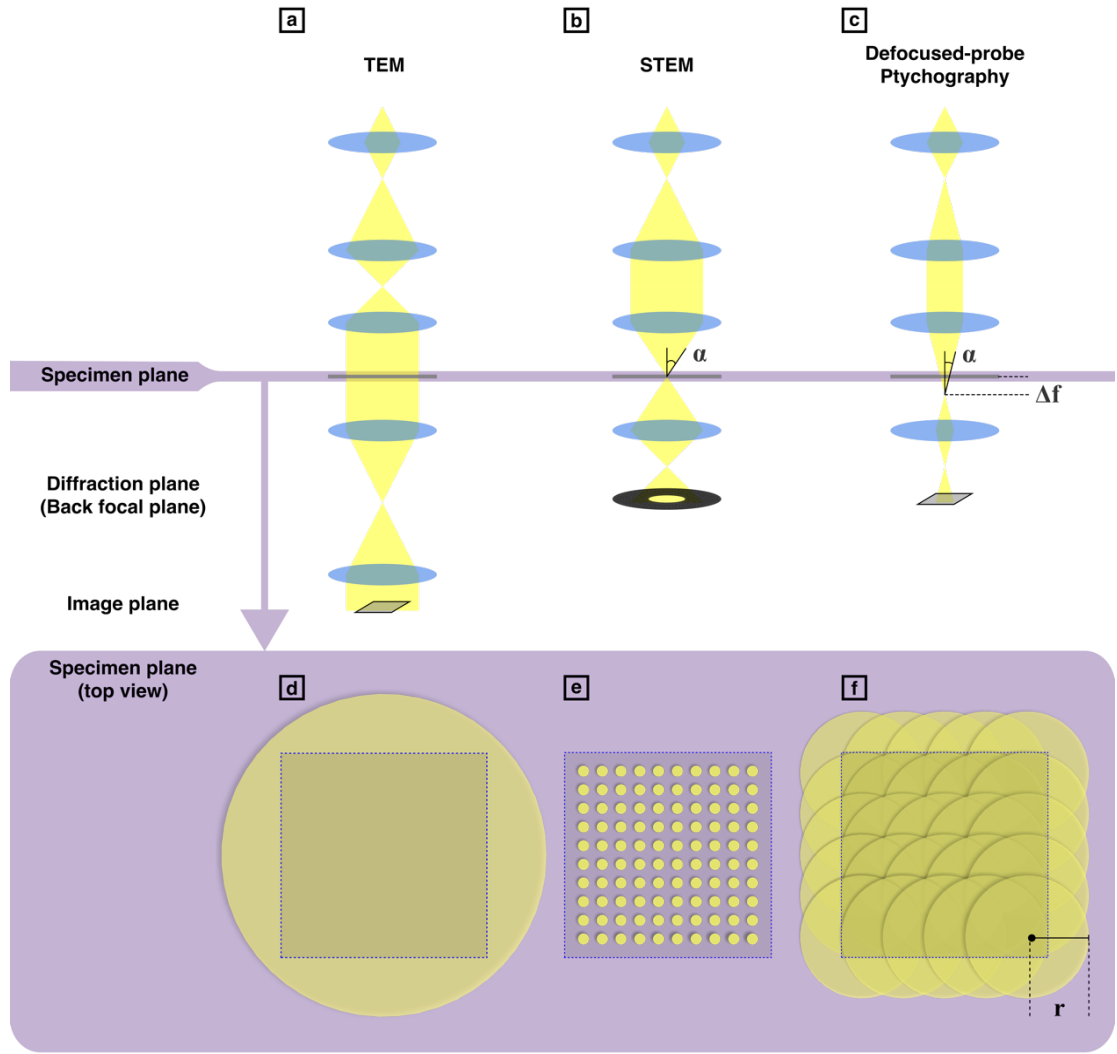

*Figure 1. Optical geometries for TEM, STEM, and defocused ptychography. Conventional TEM (a) uses parallel illumination (e) and records the spatial variations in image intensity. STEM (b) uses a converged electron probe (convergence semi-angle, α > 0°) that scans the sample. Common data acquisition in STEM collects information with a defined (usually annular) detector. For each probe position, a single intensity value (d) based on whole-detector integration is assigned to the corresponding pixel position in the image. Defocused-probe ptychography (c) uses pixelated detectors to collect an array of CBED patterns over the scanning area. A negative defocus value ($\Delta f < 0$) generates a small underfocused illumination spot ($r = \tan\alpha \cdot \Delta f$) that results in overlap between nearby probe positions (f). Top view of the specimen plane (purple) shows the fluence distribution (yellow) of the three techniques (d-f), where the dashed squares indicate the region of interest from which data is recorded.*

## Ptychography of biological specimens

Following advances in electron ptychography at a variety of fluence levels ($10^2$ - $10^6$ e⁻/Å²) using radiation resistant materials [17–21], it became apparent that these methods could also be used under the low-fluence conditions necessary for studies of biological samples.

Using simulated data Pelz *et al*. [4] explored defocused probe ptychography for varying structures (64 kDa to 4 MDa) at fluences of 5 and 20 e⁻/Å². By comparing the signal-to-noise ratio (SNR) and the Fourier ring correlation (FRC) of ptychographic reconstructions with defocused TEM ($\Delta f$ = -1.6 μm) and phase-plate TEM ($\Delta f$ = -50 nm), they demonstrated that ptychography yields high SNR at high spatial frequencies (sub-nanometre resolution). Consequently, they proposed a reduction in the number of particles needed for averaging in

SPA, and 2D resolution of 3.4-2.9 Å was recovered by averaging only tens of ptychographic phases.

Experimental results using cryo-electron ptychography from biological specimens under low-fluence conditions were first reported by Zhou *et al.* [16], for frozen hydrated rotavirus DLPs and HIV-1 VLPs at a series of fluences comparable to those used in cryo-EM (22.8 to 5.7 e$^-$/Å$^2$). This data clearly resolved the viral protein 6 (VP6) trimers on the surface of rotavirus (Figure 2).

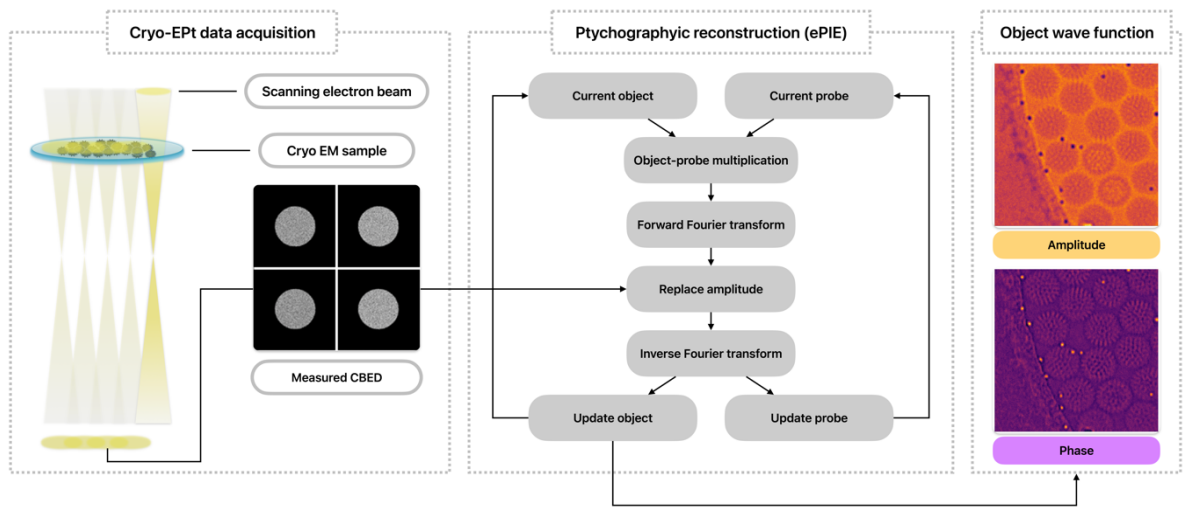

*Figure 2. Cryo-electron ptychography workflow demonstrating data acquisition and object wave function reconstruction of rotavirus at cryogenic temperature. The reconstruction was generated using the ePIE algorithm [22] on experimental data taken from [16]. $127 \times 127$ CBED patterns were used to reconstruct the object wave function with a field of view of $394\ nm\ \times\ 394\ nm$ with $0.46\ nm$ pixel size ($\Delta x$).*

Progress in SPA 3D reconstructions using reconstructed cryo-EPty phase data has been further described by Pei *et al*. [3] where frozen hydrated rotavirus DLPs were reconstructed to 1.86 nm resolution using only ~500 particles. Although this resolution is lower than that achieved using cryo-EM, the limitation arises from a combination of the low convergence angle used and an insufficient number of particles. However, in this work, a simulation study on apoferritin also showed that 2826 particles reconstructed with a 15 mrad convergence semi-angle would lead to a 3D resolution of 2.2 Å. Cryo-EPty data collected at varying convergence semi-angles were also combined to create a wide bandwidth range of spatial frequencies for information transfer of both low- and high-resolution features.

Whilst a detailed description of the complex workflows currently used in TEM SPA is outside the scope of this review, we emphasise that these can be used in an essentially unmodified form with ptychographic phase data. Moreover, as described above the advantages of using a ptychographic dataset as input arise from the tuneable information transfer which in turn produces higher contrast data requiring a lower number of classified particles for 3D reconstruction [3].

## Practical aspects of cryo-electron ptychography

**Field of view**

Defocused-probe ptychography decouples resolution from the field of view (FOV), allowing in principle, characterisation of large heterogenous biological samples at high resolution. The resolution of ptychography is theoretically diffraction-limited (*i.e.* $r_d = 0.61\lambda/\alpha$, where $\lambda$ is the electron wavelength and $\alpha$ is the convergence semi-angle of the probe). In practice for cryo-EPty the resolution is determined by the angular range in the CBED pattern which must be adequately sampled. If this is satisfied, then the field of view is simply given by the number of probe positions used and the overlap between them. In contrast, the FOV in most EM methods is typically constrained by the minimum image sampling required for a resolution of a specific feature size and the detector array size ($N_{i=1,2}$) or for scanning techniques, the scan array size ($n_{i=1,2}$), where $i$ represents one of the two-dimensions in a plane (Table 1). Ptychography, however, can utilise both $N_i$ and $n_i$ in conjunction with a computational setting of reciprocal-space sampling (Δk) to freely define the real-space sampling (Δx) of an object. As a result, the maximum FOV is primarily limited by the area accessible by the scanning system, which, as demonstrated for cryo-STEM [9], is sufficient to cover areas larger than individual cells. This large FOV in combination with high contrast and high resolution make cryo-EPty ideally suited to take advantage of newly developed 2D template matching to locate and classify proteins in native cellular environments [23,24].

**Fluence and Flux control**

The resolution of cryo-electron biological imaging is ultimately limited by electron beam damage [25–28]. Estimating the total electron fluence for the various techniques mentioned earlier is relatively straightforward (Table 1). Despite details of differences in formulation, the critical fluence for any specific specimen at high resolution will likely fall within the same order of magnitude (*e.g.* 10~100 e$^-$/Å$^2$). In contrast to conventional TEM and STEM geometries, in which the flux ($d'$) is simply the time derivative of the time integrated fluence ($d$) given by the product of fluence and dwell time, the flux of ptychography contains additional tuneable parameters, specifically the convergence semi-angle and defocus (Table 1). In ptychography, each point on the sample is exposed multiple times by overlapping probes. Given that heat and charge diffusion processes [26,29] play important roles in radiation damage and therefore affect the resolution [25], it is important to develop alternative scanning patterns that are better optimised for damage and charge control. The availability of vector scan generators designed for arbitrary (including sparse) scanning patterns in STEM [30] will in future be essential for exploring the detailed nature of radiation damage in biological specimens in vitreous ice.

*Table 1. Comparison of ptychographic methods to TEM and STEM.*

| Method | TEM | STEM | Ptychography | | |
|---|---|---|---|---|---|
| Sub types | | | Focused-probe ptychography | Defocused-probe ptychography | Fourier ptychography |
| Detector | Pixelated detector | "Integration" detector | Pixelated detector | | |
| Illumination type | Parallel | Convergent | Convergent | Convergent | Parallel (tilted) |
| Post-processing | No | No | Yes | | |
| Dataset dimensions | $N_1 \times N_2$ ($N_{i=1,2}$: detector pixel array) | $n_1 \times n_2$ ($n_{i=1,2}$: scan array) | $N_1 \times N_2 \times n_1 \times n_2$ | $N_1 \times N_2 \times n_1 \times n_2$ | $N_1 \times N_2 \times n_t$ ($n_t$: number of tilt angles) |
| Result | Phase-contrast image $N_1 \times N_2$ | ADF/ABF/BF/iDPC-STEM image $n_1 \times n_2$ | Complex object wave function | | |
| | | | $n_1 \times n_2 \times 2$ | $\frac{n_1 \Delta x_s}{\Delta x} \times \frac{n_2 \Delta x_s}{\Delta x} \times N_1 \times N_2 \times 2$ | $> N_1 \times N_2 \times 2$ |

|  |  |  |  | ($\Delta x_s$: step size; $\Delta x$: real-space sampling) |  |
| ---: | :---: | :---: | :---: | :---: | :---: |
| Data collection plane | Image plane | Diffraction plane | Diffraction plane | Diffraction plane | Image plane |
| Overlap plane | n/a |  | Diffraction plane | Sample plane | Diffraction plane |
| Fluence | $d_{TEM} = \dfrac{i \cdot t/e^-}{\Delta x^2 \cdot N_1 \cdot N_2}$ ($i$: electron beam current, $t$: exposure time) | $d_{STEM} = \dfrac{i \cdot t_{dwell}/e^-}{\Delta x_s^2} = \dfrac{i \cdot t_{dwell}/e^-}{\Delta x^2}$ ($t_{dwell}$: dwell time) |  | $d_{ptycho} = \dfrac{i \cdot t_{dwell}/e^-}{\Delta x_s^2}$ | $d_{\mathcal{F}-ptycho} = n_t \cdot d_{TEM}$ |
| Flux | $d'_{TEM} = \dfrac{i/e^-}{\Delta x^2 \cdot N_1 \cdot N_2}$ | $d'_{STEM} = \dfrac{i/e^-}{\Delta x_s^2} = \dfrac{i/e^-}{\Delta x^2}$ |  | $d'_{ptycho} = \dfrac{i/e^-}{\pi (\tan \alpha \cdot \Delta f)^2}$ $\approx \dfrac{i/e^-}{\pi (\alpha \cdot \Delta f)^2}$ ($\alpha$: convergence semi-angle, $\Delta f$: defocus) | $d'_{\mathcal{F}-ptycho} = d'_{TEM}$ |
| Achieved SPA resolution (sample, conditions) | 1.22 Å (apoferritin, 363126 particles, 40 e/Å$^2$) [31] | 3.5 Å (tobacco mosaic virus, 75616 asymmetrical units, 35 e/Å$^2$) [32] | n/a | 18.6 Å (rotavirus, 498 particles, 25 e/Å$^2$) [3] | n/a |

**Reconstruction algorithms**

Ptychographic reconstruction algorithms developed for optical and X-ray ptychography [33] can be adapted for processing electron data. These algorithms can be divided into two groups based on their updating process characteristics.

Sequential methods (the 'PIE family') are derived from the original 'ptychographic iterative engine' (PIE) algorithm [14]. The extended version of PIE, or ePIE [22] removes the requirement of a known illumination function (probe) and is more widely used for this reason (Figure 2). Subsequently modified PIE algorithms, e.g. rPIE and mPIE [34], include additional *regularisation* and *momentum*, which improve the convergence and stabilisation of the reconstruction.

An alternative family of iterative algorithms, including difference map (DM) [35] and relaxed averaged alternating reflections (RAAR) [36], are designed to process all diffraction patterns before performing object/probe updates. These algorithms are therefore sometimes known as 'batch methods', performing one global update across the object only at the end of each individual iteration. Batch methods are often simpler to implement in a highly parallelised manner, although carefully designed synchronisation of wave function updates is required.

## Emerging directions

A key challenge in cryo-EPty is to improve the resolution to a level comparable to state-to-the-art cryo-TEM. Combining multiple convergence semi-angles, either experimentally or computationally, is one strategy together with the use of a higher number of particles and the amalgamation of high resolution and high contrast data.

One promising direction would be to utilise the resolution invariant wide FOV capability of ptychography to investigate large heterogeneous structures at a resolution and sampling higher than those possible in cryo-(S)TEM. Cryo-ptychographic tomography is another area that could generate significant impact. In addition to directly applying 3D reconstruction algorithms to ptychographic reconstructions from multiple tilt angles [37,38], the intrinsic 3D information in the recovered wavefunction [39,40] could be used to potentially reduce the number of tomographic tilt angles required. Other scanning phase recovery techniques, such as focused-probe ptychography [41] and iDPC [12], have been applied under low fluence conditions to biological samples [32,42]. Fourier ptychography, which employs a similar electron optical alignment to standard cryo-TEM, has been successfully implemented in light microscopy [43], X-ray microscopy [44], and electron microscopy for radiation resistant samples at room temperature [45]. The experience gained from these implementations may provide a future roadmap for the wider adoption of this method. Hardware developments, including new phase plates [46,47] and customised apertures as have been demonstrated for TEM and which impose known phase and/or amplitude information on the probe [48], should enhance reconstruction, particularly at low fluence, by encoding known phase structure into the probe as has been demonstrated for X-Ray ptychography [49] and predicted by simulation for cryo-EPty [4]. Exploring the performance of these techniques at different accelerating voltages [17,50] will also undoubtedly provide additional future insights.

## Future perspectives

There are numerous aspects of the current cryo-EPty workflow that require improvement. Data acquisition, reconstruction and instrument calibration need to be streamlined, automated to increase throughput. Sample-related issues, such as charging, drift, and damage, also need to be eliminated. New approaches that have mitigated these issues in other cryo-TEM techniques [51–53] may also help in defocused probe ptychography.

There are several reasons to be optimistic about the future of cryo-EPty. Firstly, electron ptychography as a technique is not electron optically resolution-limited, and data from the physical sciences convincingly demonstrate that atomic resolution is possible. Secondly, the emergence of dedicated high-performance hardware is rapidly advancing. Direct electron detectors [7] will soon be sufficiently fast and efficient that every electron from the brightest electron source will be captured and individually counted [54]. Computing infrastructure and more efficient algorithm implementations are keeping pace with the increase in data generation [55]. Essential components for fluence-efficient cryo-EPty experiments, such as vector scan generators and electrostatic beam deflectors [56], can now be synchronized and integrated to ensure that electrons are delivered to any desired position on the sample, useful for wide FOV studies of heterogeneous structures. These developments together with simulations show that resolutions matching (or exceeding) those achieved using conventional cryo-EM are achievable at comparable fluence and flux. Finally, we note that the use of sparse data acquisition, inpainting and the use of machine learning to process ptychographic datasets may further reduce the required fluence.

**Conflict of interest statement**
None declared.

**Acknowledgements**
C.H., J.S.K., A.I.K. acknowledge The Rosalind Franklin Institute, funded by the UK Research and Innovation, Engineering and Physical Sciences Research Council. This research did not receive any specific grant from funding agencies in the public, commercial, or not-for-profit sectors.

# References

(Annotations of key references)
* [4]
Computational study that explores low fluence ptychographic reconstructions of varied defocused probe types for comparison of signal-to-noise ratios (SNR). The work demonstrated that ptychography can be applied to biological specimens for 2D averaged atomic resolution structure (3.4-2.9 Å) using only tens of particles due to a high SNR.

** [16]
The electron ptychographic method was used in cryogenic experiments to recover complex phase and amplitudes of frozen-hydrated rotavirus DLPs, HIV-1 VLPs and also resin embedded adenovirus infected cells at 4-27 e-/Å² total fluence values. The data produced using a defocused probe did not have contrast reversals and thus produced directly interpretable, high contrast data on particles and heterogeneous samples.

** [3]
SPA 3D reconstructions were generated using experimental ptychographic data of frozen-hydrated rotavirus DLPs for 1.86nm resolution using approximately 500 particles. A combination of data collected at varying convergence angles produced a wide spatial frequency range for information transfer of both low- and high-resolution features.

1. Vinothkumar KR, Henderson R: **Single particle electron cryomicroscopy: trends, issues and future perspective**. *Q Rev Biophys* 2016, **49**:e13.

2. Berger C, Premaraj N, Ravelli RBG, Knoops K, López-Iglesias C, Peters PJ: **Cryo-electron tomography on focused ion beam lamellae transforms structural cell biology**. *Nat Methods* 2023, **20**:499–511.

3. Pei X, Zhou L, Huang C, Boyce M, Kim JS, Liberti E, Hu Y, Sasaki T, Nellist PD, Zhang P, et al.: **Cryogenic electron ptychographic single particle analysis with wide bandwidth information transfer**. *Nat Commun* 2023, **14**:3027.

4. Pelz PM, Qiu WX, Bücker R, Kassier G, Miller RJD: **Low-dose cryo electron ptychography via non-convex Bayesian optimization**. *Sci Rep-uk* 2017, **7**:9883.

5. Chen Z, Jiang Y, Shao Y-T, Holtz ME, Odstrčil M, Guizar-Sicairos M, Hanke I, Ganschow S, Schlom DG, Muller DA: **Electron ptychography achieves atomic-resolution limits set by lattice vibrations**. *Science* 2021, **372**:826–831.

6. Nellist PD: **Scanning Transmission Electron Microscopy, Imaging and Analysis**. 2011:91–115.

7. Levin BDA: **Direct detectors and their applications in electron microscopy for materials science**. *J Phys Mater* 2021, **4**:042005.

8. Ophus C: **Four-Dimensional Scanning Transmission Electron Microscopy (4D-STEM): From Scanning Nanodiffraction to Ptychography and Beyond**. *Microsc Microanal* 2019, **25**:563–582.

9. Wolf SG, Houben L, Elbaum M: **Cryo-scanning transmission electron tomography of vitrified cells**. *Nat Methods* 2014, **11**:423–428.

10. Wolf SG, Elbaum M: **CryoSTEM tomography in biology**. *Methods Cell Biol* 2019, **152**:197–215.

11. Crewe AV, Wall J, Langmore J: **Visibility of Single Atoms**. *Science* 1970, **168**:1338–1340.

12. Lazić I, Bosch EGT, Lazar S: **Phase contrast STEM for thin samples: Integrated differential phase contrast**. *Ultramicroscopy* 2016, **160**:265–280.

13. Clausen A, Weber D, Ruzaeva K, Migunov V, Baburajan A, Bahuleyan A, Caron J, Chandra R, Halder S, Nord M, et al.: **LiberTEM: Software platform for scalable multidimensional data processing in transmission electron microscopy**. *J Open Source Softw* 2020, **5**:2006.

14. Rodenburg JM: **Ptychography and related diffractive imaging methods**. *Adv Imag Electr Phys* 2008, **150**:87 184.

15. Rodenburg JM, Faulkner HML: **A phase retrieval algorithm for shifting illumination**. *Appl Phys Lett* 2004, **85**:1 187.

16. Zhou L, Song J, Kim JS, Pei X, Huang C, Boyce M, Mendonça L, Clare D, Siebert A, Allen CS, et al.: **Low-dose phase retrieval of biological specimens using cryo-electron ptychography**. *Nature Communications* 2020, **11**.

17. Humphry MJ, Kraus B, Hurst AC, Maiden AM, Rodenburg JM: **Ptychographic electron microscopy using high-angle dark-field scattering for sub-nanometre resolution imaging**. *Nat Commun* 2012, **3**:730.

18. Putkunz CT, D'Alfonso AJ, Morgan AJ, Weyland M, Dwyer C, Bourgeois L, Etheridge J, Roberts A, Scholten RE, Nugent KA, et al.: **Atom-Scale Ptychographic Electron Diffractive Imaging of Boron Nitride Cones**. *Phys Rev Lett* 2012, **108**:073901.

19. Yang H, Rutte RN, Jones L, Simson M, Sagawa R, Ryll H, Huth M, Pennycook TJ, Green MLH, Soltau H, et al.: **Simultaneous atomic-resolution electron ptychography and Z-contrast imaging of light and heavy elements in complex nanostructures**. *Nat Commun* 2016, **7**:12532.

20. D'Alfonso AJ, Allen LJ, Sawada H, Kirkland AI: **Dose-dependent high-resolution electron ptychography**. *J Appl Phys* 2016, **119**:054302.

21. Song J, Allen CS, Gao S, Huang C, Sawada H, Pan X, Warner J, Wang P, Kirkland AI: **Atomic Resolution Defocused Electron Ptychography at Low Dose with a Fast, Direct Electron Detector**. *Sci Rep* 2019, **9**:3919.

22. Maiden AM, Rodenburg JM: **An improved ptychographical phase retrieval algorithm for diffractive imaging**. *Ultramicroscopy* 2009, **109**:1256 1262.

23. Lucas BA, Himes BA, Xue L, Grant T, Mahamid J, Grigorieff N: **Locating macromolecular assemblies in cells by 2D template matching with cisTEM**. *Elife* 2021, **10**:e68946.

24. Lucas BA, Zhang K, Loerch S, Grigorieff N: **In situ single particle classification reveals distinct 60S maturation intermediates in cells**. *eLife* 2022, **11**:e79272.

25. Egerton RF, Hayashida M, Malac M: **Transmission electron microscopy of thick polymer and biological specimens**. *Micron* 2023, **169**:103449.

26. Karuppasamy M, Nejadasl FK, Vulovic M, Koster AJ, Ravelli RBG: **Radiation damage in single-particle cryo-electron microscopy: effects of dose and dose rate**. *J Synchrotron Radiat* 2011, **18**:398 412.

27. Allegretti M, Mills DJ, McMullan G, Kühlbrandt W, Vonck J: **Atomic model of the F420-reducing [NiFe] hydrogenase by electron cryo-microscopy using a direct electron detector**. *eLife* 2014, **3**:e01963.

28. Bartesaghi A, Matthies D, Banerjee S, Merk A, Subramaniam S: **Structure of β-galactosidase at 3.2-Å resolution obtained by cryo-electron microscopy**. *Proc National Acad Sci* 2014, **111**:11709–11714.

29. Schreiber MT, Maigné A, Beleggia M, Shibata S, Wolf M: **Temporal dynamics of charge buildup in cryo-electron microscopy**. *J Struct Biology X* 2023, **7**:100081.

30. Velazco A, Béché A, Jannis D, Verbeeck J: **Reducing electron beam damage through alternative STEM scanning strategies, Part I: Experimental findings**. *Ultramicroscopy* 2022, **232**:113398.

31. Nakane T, Kotecha A, Sente A, McMullan G, Masiulis S, Brown PMGE, Grigoras IT, Malinauskaite L, Malinauskas T, Miehling J, et al.: **Single-particle cryo-EM at atomic resolution**. *Nature* 2020, doi:10.1038/s41586-020-2829-0.

32. Lazić I, Wirix M, Leidl ML, Haas F de, Mann D, Beckers M, Pechnikova EV, Müller-Caspary K, Egoavil R, Bosch EGT, et al.: **Single-particle cryo-EM structures from iDPC–STEM at near-atomic resolution**. *Nat Methods* 2022, **19**:1126–1136.

33. Pfeiffer F: **X-ray ptychography**. *Nat Photonics* 2018, **12**:9–17.

34. Maiden A, Johnson D, Li P: **Further improvements to the ptychographical iterative engine**. *Optica* 2017, **4**:736.

35. Thibault P, Dierolf M, Menzel A, Bunk O, David C, Pfeiffer F: **High-Resolution Scanning X-ray Diffraction Microscopy**. *Science* 2008, **321**:379–382.

36. Marchesini S, Krishnan H, Daurer BJ, Shapiro DA, Perciano T, Sethian JA, Maia FRNC: **SHARP: a distributed GPU-based ptychographic solver**. *J Appl Crystallogr* 2016, **49**:1245–1252.

37. Pelz PM, Griffin S, Stonemeyer S, Popple D, Devyldere H, Ercius P, Zettl A, Scott MC, Ophus C: **Solving Complex Nanostructures With Ptychographic Atomic Electron Tomography**. *Arxiv* 2022,

38. Ding Z, Gao S, Fang W, Huang C, Zhou L, Pei X, Liu X, Pan X, Fan C, Kirkland AI, et al.: **Three-dimensional electron ptychography of organic–inorganic hybrid nanostructures**. *Nat Commun* 2022, **13**:4787.

39. Li P, Maiden A: **Multi-slice ptychographic tomography**. *Sci Rep-uk* 2018, **8**:2049.

40. Gao S, Wang P, Zhang F, Martinez GT, Nellist PD, Pan X, Kirkland AI: **Electron ptychographic microscopy for three-dimensional imaging**. *Nat Commun* 2017, **8**:163.

41. Yang H, MacLaren I, Jones L, Martinez GT, Simson M, Huth M, Ryll H, Soltau H, Sagawa R, Kondo Y, et al.: **Electron ptychographic phase imaging of light elements in crystalline materials using Wigner distribution deconvolution**. *Ultramicroscopy* 2017, **180**:173 179.

42. Li X, Lazić I, Huang X, Wirix M, Wang L, Deng Y, Niu T, Wu D, Yu L, Sun F: **Imaging biological samples by integrated differential phase contrast (iDPC) STEM technique**. *J Struct Biol* 2022, **214**:107837.

43. Tian L, Liu Z, Yeh L-H, Chen M, Zhong J, Waller L: **Computational illumination for high-speed in vitro Fourier ptychographic microscopy**. *Optica* 2015, **2**:904–911.

44. Zheng G, Shen C, Jiang S, Song P, Yang C: **Concept, implementations and applications of Fourier ptychography**. *Nat Rev Phys* 2021, **3**:207–223.

45. Kirkland AI, Saxton WO, Chau K, Tsuno K, Kawasaki M: **Super-resolution by aperture synthesis: tilt series reconstruction in CTEM**. *Ultramicroscopy* 1995, **57**:355 374.

46. Schwartz O, Axelrod JJ, Campbell SL, Turnbaugh C, Glaeser RM, Müller H: **Laser phase plate for transmission electron microscopy**. *Nat Methods* 2019, **16**:1016–1020.

47. Turnbaugh C, Axelrod JJ, Campbell SL, Dioquino JY, Petrov PN, Remis J, Schwartz O, Yu Z, Cheng Y, Glaeser RM, et al.: **High-power near-concentric Fabry–Perot cavity for phase contrast electron microscopy**. *Rev Sci Instrum* 2021, **92**:053005.

48. Ophus C, Ciston J, Pierce J, Harvey TR, Chess J, McMorran BJ, Czarnik C, Rose HH, Ercius P: **Efficient linear phase contrast in scanning transmission electron microscopy with matched illumination and detector interferometry**. *Nat Commun* 2016, **7**:10719.

49. Günther B, Hehn L, Jud C, Hipp A, Dierolf M, Pfeiffer F: **Full-field structured-illumination super-resolution X-ray transmission microscopy**. *Nat Commun* 2019, **10**:2494.

50. Naydenova K, McMullan G, Peet MJ, Lee Y, Edwards PC, Chen S, Leahy E, Scotcher S, Henderson R, Russo CJ: **CryoEM at 100 keV: a demonstration and prospects**. *Iucrj* 2019, **6**:1086–1098.

51. Brilot AF, Chen JZ, Cheng A, Pan J, Harrison SC, Potter CS, Carragher B, Henderson R, Grigorieff N: **Beam-induced motion of vitrified specimen on holey carbon film**. *J Struct Biol* 2012, **177**:630 637.

52. Scheres SH: **Beam-induced motion correction for sub-megadalton cryo-EM particles**. *eLife* 2014, **3**.

53. Naydenova K, Jia P, Russo CJ: **Cryo-EM with sub–1 Å specimen movement**. *Science* 2020, **370**:223–226.

54. Schayck JP van, Zhang Y, Knoops K, Peters PJ, Ravelli RBG: **Integration of an Event-driven Timepix3 Hybrid Pixel Detector into a Cryo-EM Workflow**. *Microsc Microanal* 2022, **29**:352–363.

55. Pelz PM, Johnson I, Ophus C, Ercius P, Scott MC: **Real-Time Interactive 4D-STEM Phase-Contrast Imaging From Electron Event Representation Data: Less computation with the right representation**. *Ieee Signal Proc Mag* 2022, **39**:25–31.

56. Hashiguchi H, Yagi K, Jimbo Y, Sagawa R, Bloom RS, Reed B, Park ST, Masiel DJ, Ohnishi I: **Ways to Suppress Electron Beam Damage Using High-Speed Electron Beam Control by Electrostatic Shutter in Sample Observation and Analysis**. *Microsc Microanal* 2022, **28**:2200–2201.